# Funneling and Sculpting of Optical Waves Through Non-Magnetic Metasurfaces

NASIM MOHAMMADI ESTAKHRI[1,2,*] AND NOOSHIN M. ESTAKHRI[1,2,3]

[1]*Fowler School of Engineering, Chapman University, Orange, CA 92866 USA*
[2]*Schmid College of Science and Technology, Chapman University, Orange, CA 92866 USA*
[3]*Institute for Quantum Studies, Chapman University, Orange, CA 92866 USA*
**estakhri@chapman.edu*

**Abstract:** Efficient concentration and transport of electromagnetic energy through electromagnetically thick structures often requires resonant phenomenon and careful design considerations. Here, we introduce a realistic non-resonant approach based on electromagnetically thick self-dual metasurfaces that can funnel electromagnetic waves through subwavelength regions and without requiring magnetic materials. By satisfying the self-duality condition, the proposed metasurfaces support impedance-matched propagation and enable reflectionless energy transfer regardless of the metasurface thickness or structural details. This mechanism also allows selected control over the internal field while maintaining reasonable operational bandwidth. Metasurface elements are designed individually, and full-wave simulations confirm the predicted behavior in sample representative cases. The proposed framework provides a general strategy for robust electromagnetic energy routing and confinement, with potential impact in nonlinear optics, sensing and particle manipulation, near-field imaging, advanced absorber technologies, and wide-angle antenna systems.

## 1. Introduction

Electromagnetic (EM) wave reflection naturally occurs at interfaces between dissimilar materials or structures due to impedance mismatches caused by differences in their material properties or geometrical configurations. Consequently, a fraction of the incident EM energy is redirected toward the source, typically without substantial propagation into or interaction with the underlying medium. Depending on the application, reflection is typically considered an undesired loss mechanism, and design considerations are used to minimize or fully eliminate reflection when possible [1]-[7]. For instance, wavefront shaping with metasurfaces typically relies on the precise control of the phase and amplitude of the wave scattered by each constituent element. Analytical formulations are then used to determine the required phase and amplitude distributions, enabling the metasurface to achieve high conversion efficiency while producing the desired wavefront [5], [8]. On the other hand, in several emerging applications, solely engineering the transmitted wave can be insufficient. Precise control of the EM field distribution within and around the structure can be equally important to enable the potential concentration and funneling of EM energy into localized regions and through electromagnetically thick structures. Such capability can allow extended and enhanced light-matter interaction with potential applications in nonlinear optics, sensing, energy harvesting, and spectroscopy [9]-[13]. Consequently, implementation of metasurfaces capable of simultaneously maintaining high transmission efficiency while tailoring their internal field distribution provides a broader functionality than conventional wavefront-shaping or antireflection designs.

In this regard, symmetry principles in EM provide an attractive framework for designing metasurfaces and metamaterials with enhanced transmission and tailored field distributions. By properly engineering the geometry, arrangement, and constitutive material properties of the constituent elements, these symmetry principles can be exploited to suppress scattering and reflection under a variety of conditions [14]-[25]. In particular, EM self-duality has been employed to achieve both reflectionless transmission and wave funneling through electromagnetically thick structures under idealized conditions [18], [26], [27]. In these cases, electric ($\varepsilon_r \neq 1$) and magnetic ($\mu_r \neq 1$) materials are judiciously arranged to satisfy rotational and dual symmetries, leading to complete suppression of backscattering while directing EM energy through intentionally introduced apertures within the metastructure. However, although the self-duality condition guarantees a reflectionless response, its practical implementation is hindered by the requirement for magnetic materials, which are often unavailable or exhibit significant losses, especially at optical frequencies. Consequently, the realization of wave funneling and reflectionless transmission using exclusively non-magnetic materials remains an important challenge. A potentially promising alternative is to emulate self-dual behavior through effective electric and magnetic responses supported by appropriately engineered non-magnetic metamaterials, thereby eliminating the need for intrinsically magnetic materials [28], [29]. Nevertheless, synthesizing a sufficiently strong and broadband effective magnetic response at high frequencies remains challenging, limiting the practical realization of such self-dual configurations.

An interesting connection can be also made to Huygens' metasurfaces [5], [30] and Kerker-type scattering [31]-[33], where balanced electric and magnetic responses are exploited to suppress backscattering and achieve near-unity transmission for a variety of applications, including wavefront shaping. Depending on the design, this balance can be realized through either geometrically overlapping or distinct electric and magnetic elements [19], [32], [34]. These concepts provide a practical framework for approximating non-trivial self-dual behavior using non-magnetic resonant particles that support electric and magnetic Mie modes [35], thereby enabling reflection suppression without relying on intrinsically magnetic materials.

Building on these developments, in this work we present a new approach for realizing wave funneling and field sculpting in metasurfaces and electromagnetically thick structures using exclusively non-magnetic meta-atoms. The proposed designs consist of effective self-dual metastructures formed by core-shell nanoparticles that exhibit broadband transmission, negligible reflection, and strong concentration of EM energy within the air gaps between neighboring meta-atoms. The nanoparticles are first designed to support the desired electric and magnetic Mie responses required to satisfy the effective self-duality condition. We then investigate the performance of a single-layer metasurface and demonstrate, through full-wave simulations, that its reflectionless response enables multiple layers to be cascaded while preserving high transmission and wave-funneling characteristics. Furthermore, the proposed structures remain robust to abrupt variations between adjacent layers provided that the effective self-duality condition is maintained. Finally, we demonstrate that controlled polarization rotation within the structure can be achieved through appropriate interlayer rotations and show that metastructures composed of different combinations of electric- and magnetic-response elements continue to exhibit near-zero reflection as long as the effective self-duality condition is satisfied.

## 2. Meta-atom design and characterization

The proposed structure consists of four meta-atoms arranged in a two-dimensional periodic lattice, as shown in Figs. 1(a)-(b). For an ideal self-dual metasurface, the self-duality condition, $\mu_r(\varphi+\pi/2)=\varepsilon_r(\varphi)$, must be satisfied at every cross section along the direction of wave propagation (the -$z$-direction in this work) [18]. Here, our objective is to relax the requirement for intrinsically magnetic materials by replacing the ideal magnetic scatterers with non-magnetic particles that exhibit equivalent scattering characteristics. Assuming electromagnetically small particles, the scattering and absorption properties of each meta-atom are fully characterized by the first-order electric and magnetic Mie coefficients (or, equivalently, their electric and magnetic polarizabilities). Under these assumptions, it follows that, for a self-dual metasurface composed of spherical meta-atoms satisfying $\mu_r(\varphi+\pi/2)=\varepsilon_r(\varphi)$, the first-order electric Mie coefficient, $C_1^{\mathrm{TM}}$, of the dielectric spheres must equal the first-order magnetic Mie coefficient, $C_1^{\mathrm{TE}}$, of the magnetic spheres, following the notation in [36] and assuming an $e^{j\omega t}$ time convention. Consequently, by engineering the Mie coefficients of non-magnetic particles to satisfy the same balance, the scattering response of an ideal self-dual metasurface can be closely approximated. This approximation is expected to remain valid provided that the electric and magnetic dipolar responses remain balanced, higher-order Mie modes are negligible (or also balanced), and inter-particle coupling does not significantly perturb the scattering characteristics.

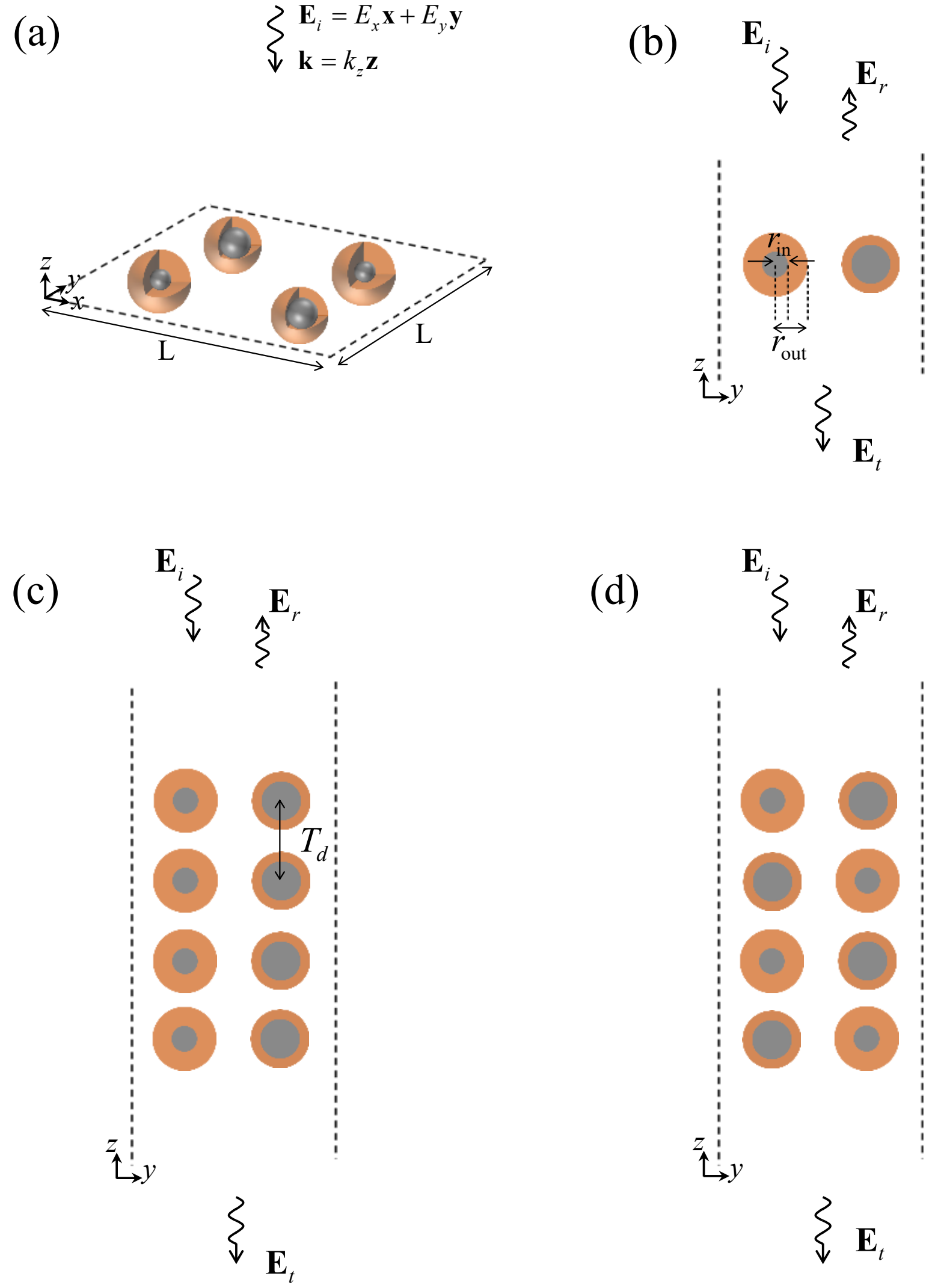

Fig. 1. (a) Unit cell of the two-dimensional metasurface, consisting of four meta-atoms and periodically repeated along the $x$- and $y$-directions with period $L$. Each meta-atom is a core-shell nanoparticle, where the orange and gray regions represent silicon and silver, respectively. (b) Side view of the metasurface shown in (a) in the $yz$-plane. (c) Four-layer metastructure composed of identical metasurface layers. (d) Four-layer metastructure with alternating rotated layers. In all panels, the dashed lines denote periodic boundary conditions.

The metasurface is illuminated at normal incidence by a linearly polarized plane wave, $\mathbf{E}_i = E_x\mathbf{x} + E_y\mathbf{y}$ , with a design wavelength of 1550 nm. The unit-cell periodicity is chosen to satisfy the single Floquet mode condition, i.e., $L < 1550\,\mathrm{nm}$, thereby ensuring that higher-order diffraction modes are evanescent. The four core-shell nanoparticles are positioned at the centers of the four quadrants of the unit cell, with the particles located along each diagonal being identical. All meta-atoms consist of a silicon shell [37], surrounding a silver core [38]. The inner and outer radii of each particle are optimized using genetic algorithm to realize the desired electric and magnetic scattering coefficients (i.e., $C_1^{\mathrm{TM}}$ and $C_1^{\mathrm{TE}}$), as discussed in the following. Although a simple optimization strategy is adopted here, the proposed design framework is readily compatible with more advanced inverse-design techniques, including topology optimization or machine-learning-based approaches [39]-[43], which could enable more complex self-dual metastructures with enhanced performance metrics (e.g., bandwidth, localization, etc.).

Two sets of meta-atoms are designed based on the effective self-duality condition discussed in the previous section. In both cases, the particles are engineered such that the first-order electric Mie coefficient, $C_1^{\mathrm{TM}}$, of the first meta-atom equals the first-order magnetic Mie coefficient, $C_1^{\mathrm{TE}}$, of the second meta-atom and vice versa while higher order scattering coefficients are kept negligible. Consequently, the dipolar scattering response of the meta-atoms closely reproduces that of an ideal self-dual structure.

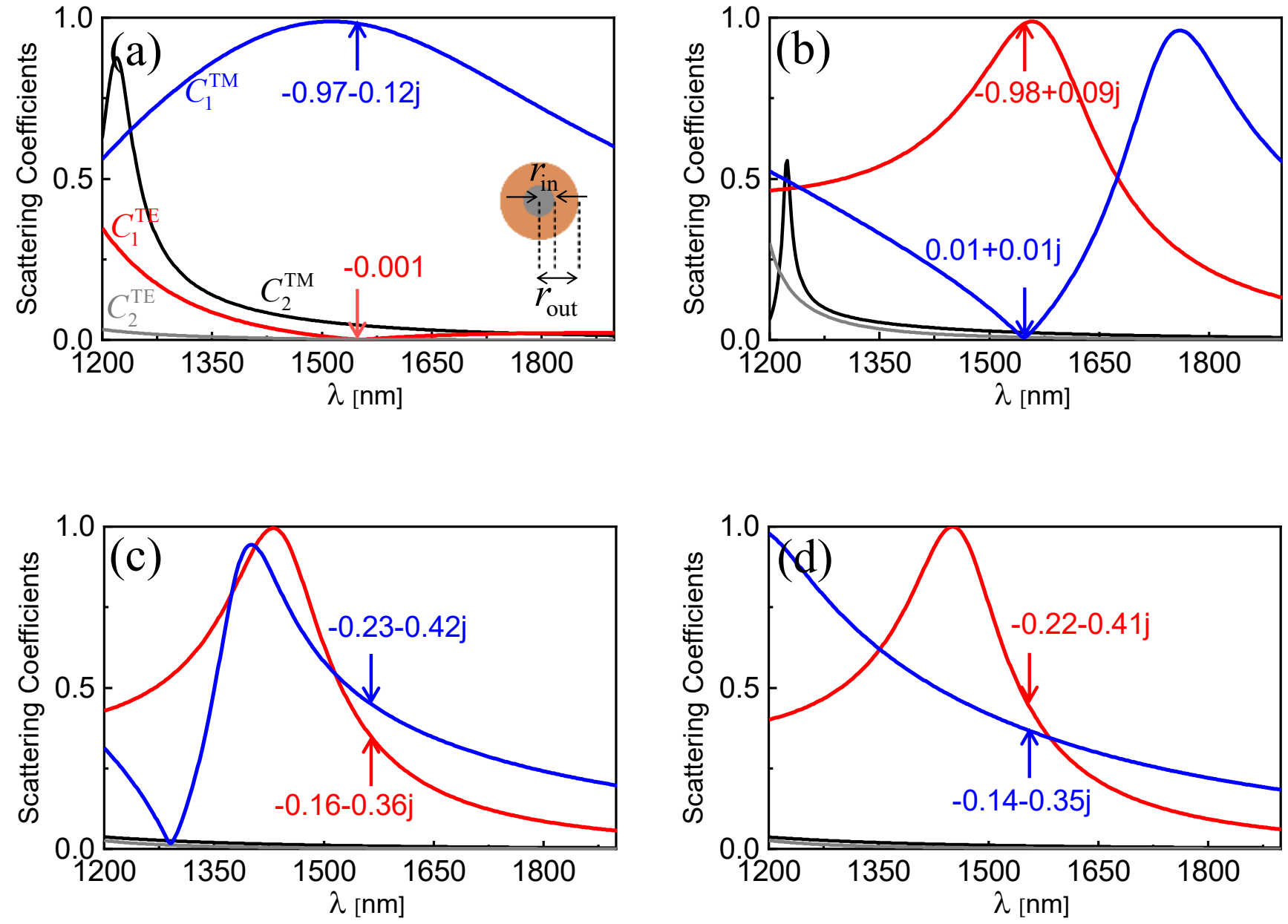

Fig. 2. Dispersion characteristics of the four designed core-shell nanoparticles. The particles shown in (a) and (b) are used in metasurface 1 (Section 3.1), while those shown in (c) and (d) are used in metasurface 2 (Section 3.2). Schematic representations of the core-shell nanoparticles are shown in the inset of (a). The color-coded curves indicate the magnitude of the first four Mie scattering coefficients, as shown. The complex values of the first two scattering coefficients are shown at a wavelength of 1550 nm in all panels. The radii of the core-shell nanoparticles are $r_{\mathrm{in}}, r_{\mathrm{out}} =$ (a) $190\,\mathrm{nm}, 253\,\mathrm{nm}$, (b) $98\,\mathrm{nm}, 234\,\mathrm{nm}$, (c) $55\,\mathrm{nm}, 200\,\mathrm{nm}$, and (d) $30\,\mathrm{nm}, 200\,\mathrm{nm}$.

For the first design, the optimization enforces, $|C_1^{\mathrm{TM}}| = 1$ and $|C_1^{\mathrm{TE}}| = 0$ for the first meta-atom, whereas $|C_1^{\mathrm{TM}}| = 0$ and $|C_1^{\mathrm{TE}}| = 1$ are prescribed for the second meta-atom. The constituent materials are fixed throughout the optimization, and candidate designs with a core radius smaller than 20 nm or an overall particle radius larger than 600 nm are discarded. The resulting optimized particles are shown in Fig. 2(a) and (b), corresponding to meta-atoms 1 and 2 of the first metasurface design, respectively. Particle dimensions are provided in the caption. At the design wavelength of 1550 nm, meta-atom 1 behaves as an almost purely electric dipole, whereas meta-atom 2 exhibits an almost purely magnetic dipole response. These particles are employed in the first metasurface design, which is characterized in Section 3.1. From the perspective of scattering, this design can be also interpreted as a spatially distributed realization of the Kerker condition [31].

For the second design, our objective is to demonstrate that effective self-duality can also be achieved using meta-atoms with mixed electric and magnetic responses, while wave funneling might be reduced in this case. To this end, meta-atom 1 is fixed as a core-shell nanoparticle with an inner radius of 30 nm and an outer radius of 200 nm. The corresponding complex first-order Mie scattering coefficients are first calculated, as shown in Fig. 2(c). The optimization of meta-atom 2 then enforces the complementary (swapped) values of these complex coefficients while maintaining negligible higher-order scattering contributions. The resulting optimized particle is shown in Fig. 2(d). Together, the particles in Fig. 2(c) and (d) constitute meta-atoms 1 and 2 of the second metasurface design, respectively. At the design wavelength of 1550 nm, both particles exhibit simultaneous electric and magnetic dipolar responses while satisfying the effective self-duality condition. These meta-atoms are employed in the second metasurface design, which is characterized in Section 3.2. Unlike the first design, which approximates a spatially distributed Kerker configuration using nearly pure electric and magnetic dipoles, this implementation more closely resembles the conventional Kerker condition, where balanced electric and magnetic responses coexist within each scattering element.

Before reporting the results for metasurfaces, we note that the dispersive behavior of the proposed meta-atoms, characterized through their Mie scattering coefficients in Fig. 2, is expected to play an important role in determining the overall performance of each metasurface. In particular, the spectral overlap and relative balance between the electric and magnetic dipolar coefficients directly influence the achievable operational bandwidth and the robustness of the effective self-duality condition. Moreover, the relative distribution of electric and magnetic responses among the meta-atoms within a given metasurface is expected to affect the strength of wave funneling and the spatial redistribution of EM energy. Specifically, a more spatially separated implementation of complementary electric and magnetic responses (e.g., in metasurface 1), is anticipated to support stronger field localization compared to configurations where both responses coexist within individual meta-atoms (e.g., in metasurface 2).

## 3. Metasurface characterization and performance

The EM response of the proposed metasurfaces is investigated through full-wave simulations [44]. Besides evaluating their reflection and transmission characteristics, we examine their ability to redistribute and funnel EM energy through the metastructure. The two metasurface configurations introduced in Section 2 are analyzed independently to illustrate the impact of the effective self-duality condition for both spatially distributed and mixed electric-magnetic scattering responses.

### *3.1 Pure electric and magnetic meta-atoms*

We begin by considering a single layer metasurface composed of the two meta-atoms characterized in Fig. 2(a) and (b). The particles are arranged in a two-dimensional square lattice with a period of $L = 1100\,\text{nm}$ and are illuminated at normal incidence by an $x$-polarized plane wave $\mathbf{E}_i = E_x\mathbf{x}$ . Full-wave simulations are performed to evaluate the reflection and transmission characteristics over the wavelength range from 1420 nm to 1680 nm. Owing to the effective self-duality condition, high transmission is expected at the design wavelength of 1550 nm.

The simulated transmission and reflection spectra are shown in Fig. 3(a). At the design wavelength, the metasurface exhibits more than 91% transmission, while less than 5% of the incident power is reflected. The remaining power is absorbed by the silver cores of the nanoparticles. Furthermore, the structure maintains a transmission exceeding 50% over the wavelength range of 1460-1650 nm, corresponding to a 190 nm half-power bandwidth. The minimum reflection occurs near 1600 nm, in agreement with the scattering characteristics of the constituent meta-atoms shown in Fig. 2(a) and (b). It is also observed that the operational bandwidth is primarily limited by the bandwidth of the magnetic dipolar resonance. Broader bandwidths could be achieved through more sophisticated particle geometries and advanced optimization techniques.

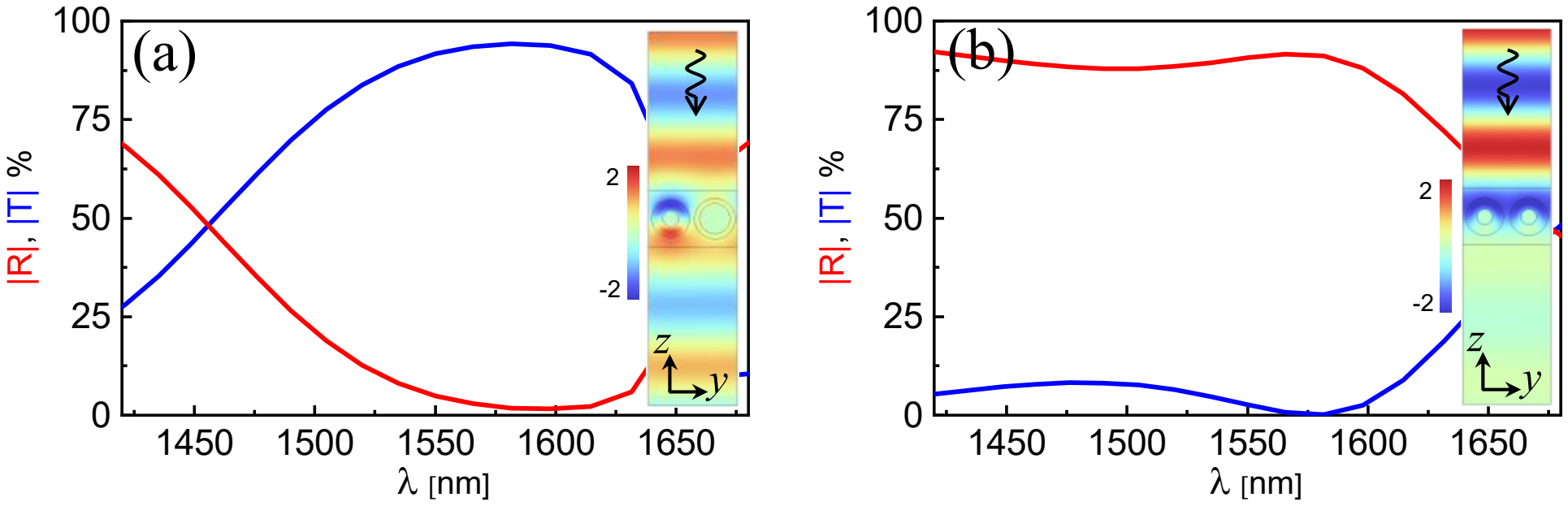

Fig. 3. (a) Simulated transmission and reflection spectra of the proposed effective self-dual metasurface composed of the meta-atoms shown in Fig. 2(a) and (b). The metasurface exhibits high transmission and suppressed reflection around the design wavelength of 1550 nm. (b) Corresponding spectra for a reference metasurface with the same lattice periodicity but composed exclusively of the meta-atom shown in Fig. 2(b). The absence of the complementary scattering response results in strong backscattering and significantly reduced transmission. The cross-polarized reflection and transmission coefficients are zero in both cases. Snapshots of the normalized $x$-component of the electric field are shown in the inset of each panel, where the fields are normalized to the amplitude of the incident electric field.

For comparison, Fig. 3(b) presents the corresponding response of a metasurface with the same lattice periodicity but composed exclusively of the meta-atom characterized in Fig. 2(b). In contrast to the proposed effective self-dual design, approximately 90% of the incident power is reflected over the considered wavelength range, with only a small fraction being transmitted. This comparison highlights the critical role of the effective self-duality condition in suppressing backscattering and enabling high transmission. The $x$-component of the electric field is also plotted in the $yz$-plane at $x = L/4$ in inset of Fig. 3, visualizing the field distribution around the particles at 1550 nm.

Given the broadband operation of the proposed metasurface and the stable angular response previously reported for ideal self-dual structures [18], a similarly robust and non-resonant response is expected from the effective self-dual implementation. This behavior is confirmed in Fig. 4(a), where the simulated transmission and reflection coefficients are plotted as a function of the angle of incidence for angles between 0 degrees and 24 degrees. The maximum angle is selected to ensure that only the fundamental Floquet mode is supported. As shown, the response remains remarkably stable over the entire angular range, with the reflected power remaining below 5%. The inset shows a snapshot of the normalized $x$-component of the electric field in the $yz$-plane (the plane of incidence) at $x = L/4$ for an incidence angle of 16 degrees, demonstrating that the wave-funneling behavior is preserved under oblique illumination.

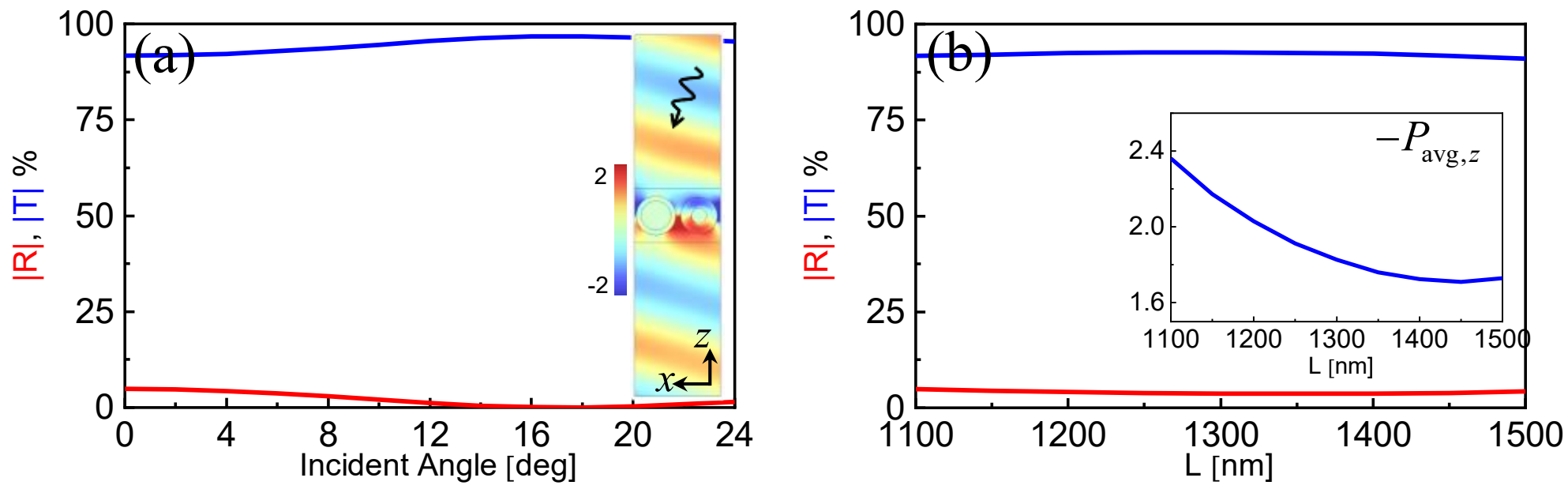

Fig. 4. (a) (a) Simulated transmission and reflection coefficients as a function of the angle of incidence. The inset shows a snapshot of the normalized *x*-component of the electric field in the *yz*-plane at $x = L/4$ for an incidence angle of 16 degrees. (b) Simulated transmission and reflection coefficients as a function of the lattice period L. The inset shows the normalized EM power density in the -z-direction at the center of the metasurface layer as a function of L.

The weak dependence on the angle of incidence suggests that the metasurface response is not dominated by narrow lattice resonances. To further verify this observation, the lattice period is varied from 1100 nm to 1500 nm while maintaining the single-Floquet-mode condition. As the lattice period increases, the coupling between neighboring particles is expected to decrease. Nevertheless, the simulated transmission and reflection coefficients remain nearly unchanged throughout the investigated range, as shown in Fig. 4(b), demonstrating that the effective self-duality condition is robust against moderate variations in the lattice periodicity. The inset of Fig. 4(b) plots the EM power density at the center of the metasurface layer as a function of the lattice period. As expected, the power density increases as the air gaps between neighboring particles become smaller, resulting in stronger field confinement. Although the minimum gap size is limited here by the spherical particle geometry, stronger localization is anticipated for alternative particle shapes, such as cubes, that permit smaller separations between adjacent meta-atoms.

After we establish the performance of a single metasurface layer, we now investigate cascaded configurations with electrically thick overall dimensions, as illustrated in Fig. 1(c) and (d) for a sample case of four cascaded layers. We first consider a stack of identical metasurface layers as in Fig. 1(c). Fig. 5(a) shows the simulated transmission and reflection coefficients as a function of the number of cascaded layers. Remarkably, the reflected power remains below 5% even as the overall thickness increases, closely resembling the behavior of ideal self-dual structures. In contrast, the transmitted power gradually decreases due to the cumulative absorption introduced by the silver cores in successive layers. As previously discussed for ideal self-dual structures, this behavior suggests a practical route toward broadband reflectionless absorbers [45], [46].

An important consideration for multilayer implementations is the effect of coupling between cascaded layers. To investigate this aspect, the separation between neighboring metasurface layers, i.e., $T_d$ , is varied in a five-layer configuration. The corresponding transmission and reflection coefficients are shown in Fig. 5(b). Despite significant variations in the interlayer spacing, the reflection remains negligible, demonstrating that the effective self-duality condition is preserved in the presence of varying interlayer couplings. This result further indicates that additional metasurface layers may be incorporated and there is considerable flexibility in the relative positioning of different layers without introducing unwanted backscattering.

Next, the EM field distributions within the multilayer structures are examined to verify wave funneling and sculpting. Fig. 5(c) illustrates the normalized power density in the propagation direction, superimposed with the electric field vectors, plotted at the center of the third metasurface layer. Here, the incident field is polarized along the $x$-direction, i.e., $\mathbf{E}_i = E_x\mathbf{x}$. As illustrated, the EM power is concentrated primarily within the air gaps separating the neighboring meta-atoms. The electric field is primarily along the oblique directions inside these regions, confirming efficient wave funneling. Interestingly, when the incident polarization is changed to $\mathbf{E}_i = E_0\mathbf{x} - E_0\mathbf{y}$, the power and field distributions change substantially. Because this polarization matches the field distribution supported by only one family of air gaps, the transmitted power is preferentially directed through those channels. The resulting side view of the power distribution is shown in Fig. 5(g), illustrating the selective concentration and transport of EM energy through the metastructure. The possibility of the localization of the EM wave over extended depths inside the metastruture suggests applications in nonlinear and quantum optics where metasurfaces have demonstrated improvements beyond traditional optical elements [47]-[51].

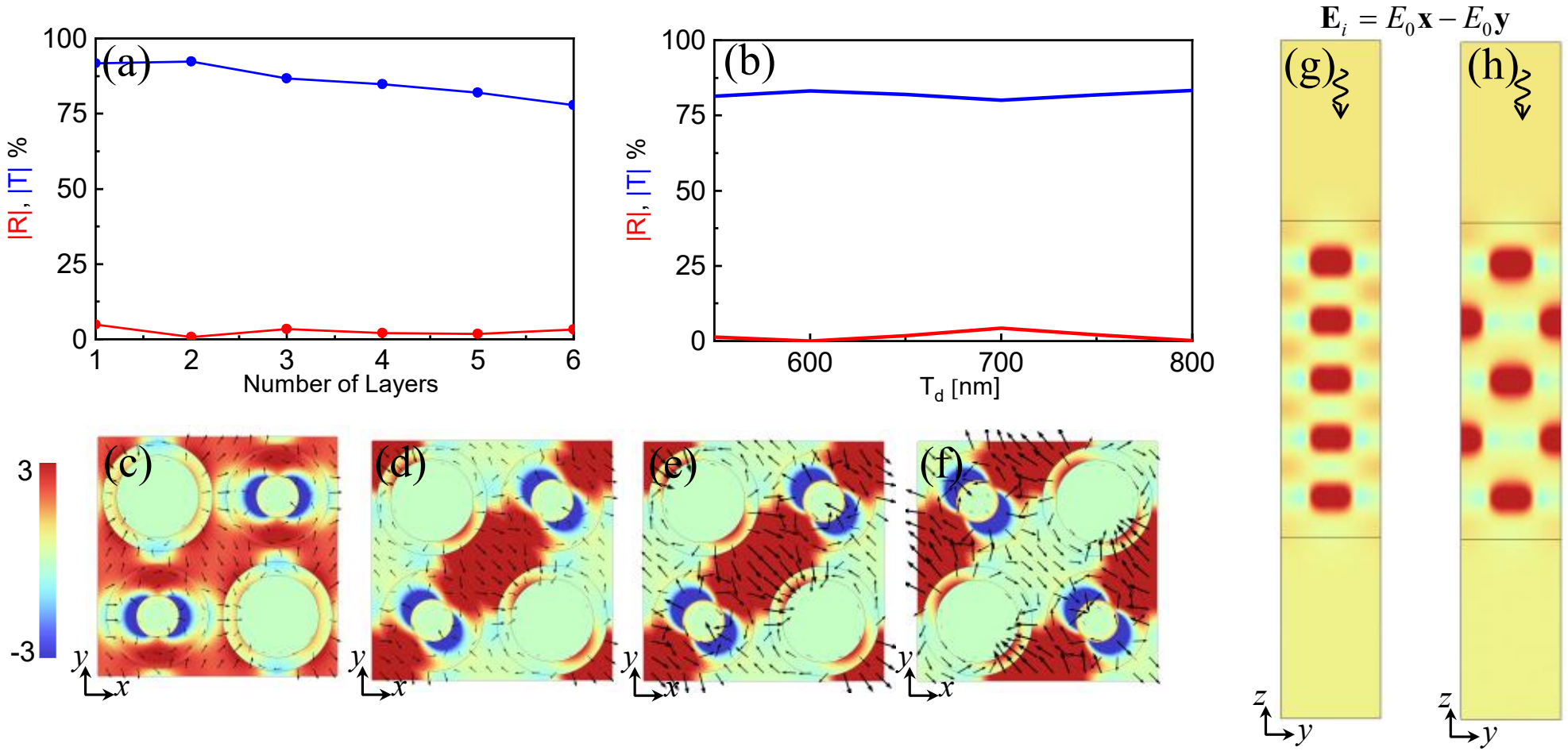

Fig. 5. (a) Simulated transmission and reflection as a function of the number of cascaded metasurface layers. (b) Simulated transmission and reflection coefficients of a five-layer metastructure as a function of the interlayer separation. Normalized power density in the propagation direction overlaid with the in plane electric field vectors at the center of the third layer of metasurface for linearly polarized incident fields with (c) $\mathbf{E}_i = E_x\mathbf{x}$, and (d) $\mathbf{E}_i = E_0\mathbf{x} - E_0\mathbf{y}$. Normalized power density in the propagation direction overlaid with the in plane electric field vectors at the center of the (e) third and (f) second layer respectively, for five-layer rotated multilayer configuration similar to in Fig. 1(d). The side view of power distribution between different layers for (g) five-layer metastructure and (h) five-layer metastructure with alternating rotated layers. All plots share same color bars as shown in panel (c).

Finally, we exploit the fact that the self-duality condition does not require the structure to remain invariant along the propagation path, provided that each transverse cross section individually satisfies the self-duality condition [18]. To investigate and demonstrate this capability, a five-layer metastructure is constructed with alternating layers rotated by 90 degrees, as illustrated in Fig. 1(d). The corresponding field distributions are shown in Fig. 5(e) and (f), where the power density and electric field vectors are plotted at the centers of the third and second layers, respectively. Although the structure and the preferred funneling gaps change abruptly from one layer to the next, the overall reflection remains negligible. Instead, EM energy follows the locally prescribed field distribution within each layer, enabling the redistribution of power between neighboring air gaps without producing any reflections. This behavior is further illustrated by the side view in Fig. 5(h), where the EM energy is observed to propagate through different air gaps. These results demonstrate the ability of the proposed effective self-dual metastructures to achieve reflectionless wave funneling together with controlled EM field sculpting throughout electrically thick structures.

The presented results in this section demonstrate that the effective self-dual metasurface successfully reproduces the key features of its ideal counterpart while relying exclusively on non-magnetic meta-atoms. The structure enables wave funneling and controlled EM field distributions that remain robust under oblique illumination, variations in lattice periodicity, and cascaded multilayer implementations. The response is also broadband. In the following section, we investigate another self-dual metasurface where the electric and magnetic response are not fully spatially separated, yet the meta-atoms support simultaneous electric and magnetic scattering responses with different strengths.

### *3.2 Mixed electric and magnetic meta-atoms*

Having demonstrated the effectiveness of the metasurfaces with spatially separated electric and magnetic dipoles, we now briefly consider the second metasurface design composed of meta-atoms exhibiting simultaneous semi-balanced electric and magnetic dipolar responses shown in Figs. 2 (c) and (d). This implementation represents a closer approximation to the conventional Kerker condition [31] and serves to illustrate that wave funneling and sculpting capabilities of these structures can be tuned by the proper choice of the meta-atoms. The particles are again arranged in a two-dimensional square lattice with a period of $L = 1100\,\mathrm{nm}$ and are illuminated at normal incidence by linearly polarized plane waves. Full-wave simulations are performed to evaluate the reflection and transmission characteristics over the wavelength range from 1200 nm to 3000 nm. Owing to the effective self-duality condition and low scattering of both particles, high transmission is expected at the design wavelength of 1550 nm and over an extended wavelength range, as illustrated in Fig. 6(a).

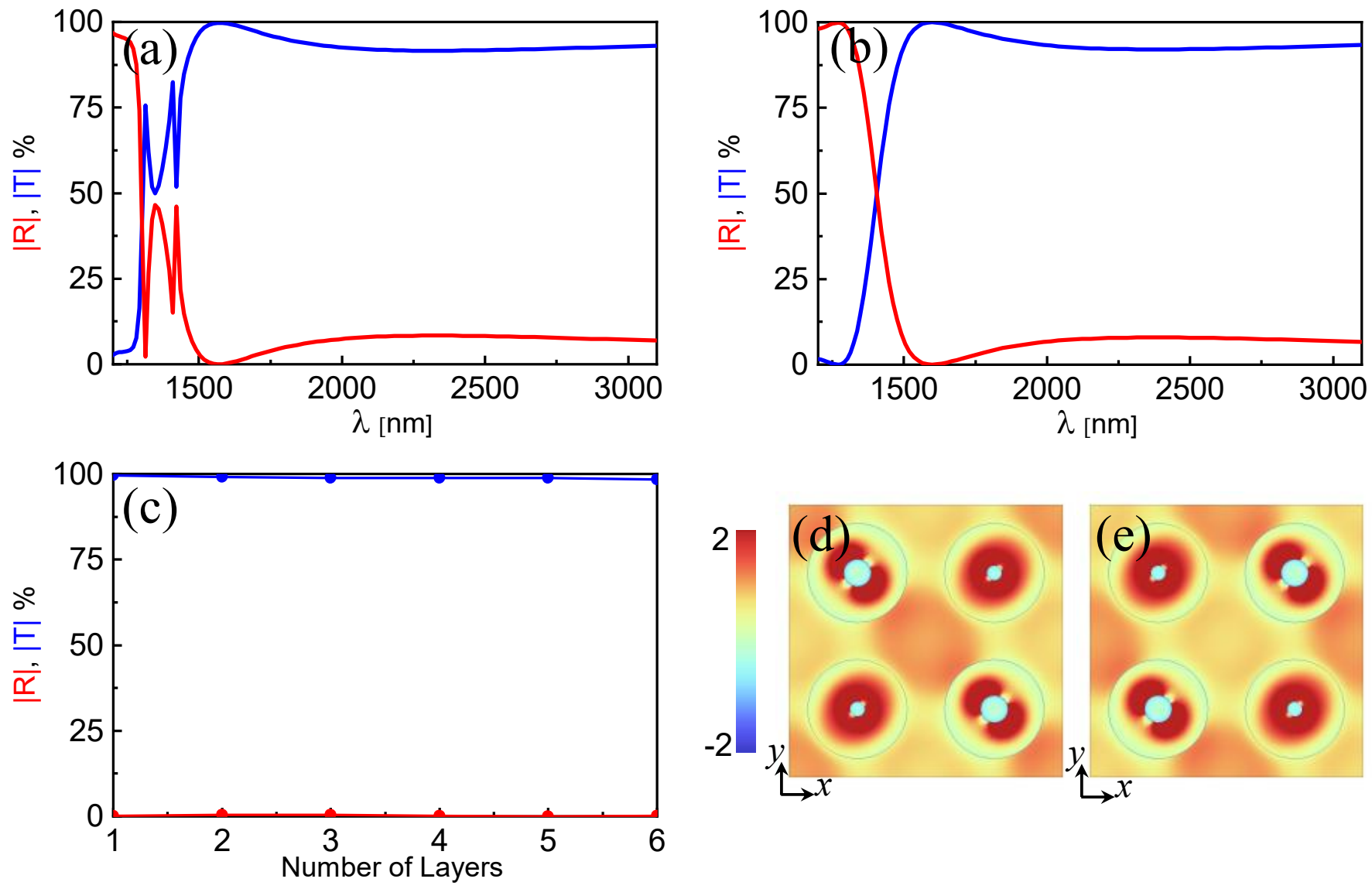

Fig. 6. (a) Simulated transmission and reflection spectra of the proposed effective self-dual metasurface composed of the meta-atoms shown in Fig. 2(c) and (d). The metasurface exhibits high transmission and suppressed reflection around the design wavelength of 1550 nm and across a broad bandwidth. (b) Corresponding spectra for a reference metasurface with the same lattice periodicity but composed exclusively of the meta-atom shown in Fig. 2(d). Due to the satisfaction of approximate self-duality in all elements, this metasurface also exhibits broadband low reflection. (c) Simulated transmission and reflection as a function of the number of cascaded metasurface layers. Normalized power density in the propagation direction

at the center of the (d) third layer and (e) second layer of the metasurface for an incident field $\mathbf{E}_i = E_0\mathbf{x} - E_0\mathbf{y}$ . Both plots share same color bars as shown in panel (d).

For comparison, Fig. 6(b) presents the corresponding response of a similar metasurface with composed exclusively of the meta-atom characterized in Fig. 2(d). Interestingly, and since both meta-atoms are approximate Kerker particles, similar broadband low reflection is maintained. Indeed, in such cases enforcing exact self-duality at single wavelength is not expected to increase the bandwidth. The dependence of transmission and reflection on the number of cascaded layers is also shown in Fig. 6(c), confirming that the low-reflection behavior is entirely preserved under multilayer operation. Finally, Figs. 6 (d) and (e) illustrates representative power distributions, highlighting the absence of obvious field concentration within the air gaps, as meta-atoms are essentially operating as matched elements.

Power distributions further confirm that the EM energy remains more uniformly distributed across the metasurface, with small preferential concentration within the air gaps. This behavior is expected, as discussed before, since both meta-atoms possess comparable electric and magnetic responses, producing homogeneous scattering. Consequently, while the effective self-duality condition still ensures high transmission and suppressed reflection, it does not, by itself, guarantee wave funneling or localized field enhancement. Instead, the degree of field localization is primarily determined by the spatial distribution of the complementary electric and magnetic scattering responses within the metastructure.

## 4. Summary and conclusions

In summary, a practical framework for realizing electromagnetically thick self-dual metastructures using exclusively non-magnetic meta-atoms is presented, with a focus on wave funneling and field sculpting. By engineering the electric and magnetic scattering responses of individual meta-atoms, the proposed designs achieve broadband transmission with minimal reflection while enabling controlled redistribution of EM energy within the structure. The results further demonstrate that wave funneling and field sculpting depend on the spatial arrangement of complementary scattering responses, providing additional flexibility for tailoring the internal EM field distribution. We envision potential applications in nonlinear optics, sensing, and absorber technologies, together with several opportunities for future developments. In particular, more sophisticated design approaches may be employed to optimize the meta-atoms for experimental realization and improved control of the air gaps [39]-[43]. Furthermore, given the broad operational bandwidth, engineering the temporal characteristics of the incident wave may enable dynamic control of the funneling effect and enhanced memory effects using metasurface configurations [53], [54].

**Disclosures.** The authors declare no conflicts of interest.

**Data availability.** Data underlying the results presented in this paper are not publicly available at this time but may be obtained from the authors upon reasonable request.